\documentclass[useAMS,usenatbib]{mnras}
\usepackage{float}
\input{boxed1.sty}
\floatstyle{simplerule}
\restylefloat{figure}
\restylefloat{table}

\usepackage{amssymb}
\usepackage{graphicx}
\usepackage{longtable}
\usepackage[table]{xcolor}
\hypersetup{
   colorlinks=true,             
   linkcolor=blue,              
   urlcolor=blue,               
   anchorcolor=blue,            
   citecolor=blue,              
}
\def\sun{\hbox{$\odot$}}
\definecolor{grey80}{rgb}{0.90,0.90,0.90}
\def\sun{\hbox{$\odot$}}

\usepackage{amsmath}

\newcounter{mycount}

\title[Dynamical environment of asteroid 21 Lutetia]{\centering \bf The dynamical environment of asteroid 21 Lutetia \\ according to different internal models}
\author[Aljbaae et al.]
{S. Aljbaae$^{1,2}$\thanks{\href{safwan.aljbaae@obspm.fr}{safwan.aljbaae@obspm.fr}}, T. G. G. Chanut$^{1,3}$, V. Carruba$^{1}$, J. Souchay$^{2}$, A. F. B. A. Prado$^{3}$ \and and A. Amarante$^{1}$\\
$^{1}$Univ. Estadual Paulista - UNESP, Grupo de Din\^amica Orbital \& Planetologia\,, Guaratinguet\'a\,, CEP 12516-410\,, SP\,, Brazil \\
$^{2}$Observatoire de Paris, SYRTE/CNRS UMR8630, 61 avenue de l'Observatoire, 75014 Paris, France\\
$^{3}$Division of Space Mechanics and Control, INPE, C.P. 515, 12227-310 S\~ao Jos\'e dos Campos, SP, Brazil}

\begin{document}

\date{Accepted ... . Received 2016 ...; in original form 2016 September 13.}

\pagerange{\pageref{firstpage}--\pageref{lastpage}} \pubyear{2002}

\maketitle
\label{firstpage}
\begin{abstract}
   One of the most accurate models currently used to represent the gravity field of irregular bodies is the polyhedral approach. In this model, the mass of the body is assumed to be homogeneous, which may not be true for a real object.  The main goal of the present paper is to study the dynamical effects induced by three different internal structures (uniform, three- and four-layers) of asteroid (21) Lutetia, an object that recent results from space probe suggest being at least partially differentiated.  The Mascon gravity approach used in the present work, consists of dividing each tetrahedron into eight parts to calculate the gravitational field around the asteroid.
The zero-velocity curves show that the greatest displacement of the equilibrium points occurs in the position of the E4 point for the four-layers structure and the smallest one occurs in the position of the E3 point for the three-layers structure.  Moreover, stability against impact shows that the planar limit gets slightly closer to the body with the four-layered structure.

We then investigated the stability of orbital motion in the equatorial plane of (21) Lutetia and propose numerical stability criteria to map the region of stable motions.  Layered structures could stabilize orbits that were unstable in the homogeneous model. 
\end{abstract}
\section{Introduction} \label{Introduction}

The main challenge for the navigators of space missions to small irregular
bodies is to derive pre-mission plans for the control of the orbits.
A lot of studies have already been focused on this issue
\citep{Scheeres_1994,Scheeres_1998a,Scheeres_1998b,Rossi_1999,Hu_2002}.
Generally, the potential of an asteroid can be estimated from its shape
assuming a homogeneous density distribution. Yet, it remains an approximation
to reality, since real bodies are affected by density irregularities.
Therefore, it seems worthwhile to discuss the effects of different mass
distributions of objects on their gravity field and, consequently,
on their orbital environment. For instance, several studies modeled the
gravitational forces of Ceres and Vesta by a spherical harmonic expansion
assuming diverse scenarios for interior structure 
\citep{Tricarico_2010,Konopliv_2011,Konopliv_2014,Park_2014}. 
In addition, the polyhedral
approach \citep{Werner_Scheeres_1997} seems more appropriate for evaluating the
gravitational forces close to the surface.  The main problem of these
approaches is the heavy computation time of the integrations. This issue
has been recently reported in \citet{Chanut_2015a} in developing a new
approach that models the external gravitational field of irregular bodies
through mascons. The authors applied the mascon gravity framework using a
shaped polyhedral source, dividing each tetrahedron into up to three parts.
That drives the attention to the possibility of taking into consideration
the structure of layers in the gravitational potential computation. 

The asteroid (21) Lutetia belongs to the main belt, the orbital space
between Mars and Jupiter. An analysis of its surface composition and
temperature, \citet{Coradini_2011} showed that Lutetia was likely formed
during the very early phases of the Solar System.  Moreover,
measurements by the European Space Agency's Rosetta have found that this
asteroid was unusually dense for an asteroid ($3.4 ~ g \cdot cm^{-3}$).
Its large density suggests that the asteroid might be a
partially differentiated body, with a dense metal-rich core \citep{Patzold_2011,Weiss_2012}. For these reasons, (21) Lutetia represents a suitable object
to test the effects of the layers structure on the gravity field.

Thus, this paper aims at computing the gravitational field associated with 
  asteroid (21) Lutetia, considering a model with different density layers.
  Moreover, we mapped the orbital dynamics of a probe-target close to it,
  taking into account this in-homogeneous model. For these purposes,
  first the physical properties of the polyhedral shape of (21) Lutetia are
  presented in section \ref{polyhedral_shape}.  Then, two models with
  different internal structures (three- and four-layers) are discussed in
  section \ref{Internal_structure}.  Moreover, the dynamical properties in
  the vicinity of our target are studied in section \ref{Dynamical_properties}.
  Here we calculated the Jacobi integral and obtained the
  zero-velocity surfaces and the particular solutions of the system. A
  numerical analysis of the stability of motions in the equatorial plane is
  presented in section \ref{Orbital_stability}. Finally, the main results of
  our study are given in section \ref{Conclusion}.

  \section{Physical properties from the polyhedral shape of Lutetia with
    uniform density}
  \label{polyhedral_shape}

The relatively large asteroid (21) Lutetia is a primordial object,
located in the inner part of the main-belt, with a perihelion of 2.036 AU
and an aphelion distance of 2.834 AU. Its eccentricity (0.164) is moderate,
and its inclination with respect to the ecliptical plane is quite small
(3.0648$^{\circ}$) \citep{Schulz_2010}. The asteroid was encountered by Rosetta
spacecraft on its way to its final target
(the comet 67P/Churyumov-Gerasimenko), at a distance of
$3168 \pm 7.5$ $km$ and a relative fly-by velocity of 14.99 $km.s^{-1}$.
The asteroid's mass was estimated  by the
gravitational field distortion of the flyby trajectory measured by the
Doppler shift of the radio signals from Rosetta
as $(1.7 \pm 0.017)\times 10^{18} ~kg$,. It is lower than the
previous estimation of $(2.59 \pm 0.24)\times 10^{18} ~kg$
obtained from asteroid to
asteroid perturbations \citep{Patzold_2011}.
Its bulk density of $3.4 \pm 0.3 ~ g.cm^{-3}$ was calculated using the volume 
determined by the Rosetta Optical, Spectroscopic, and Infrared Remote
Imaging System (OSIRIS) camera. This density is close to the density of
M-type asteroids like (216) Kleopatra \citep{Descamps_2011}. 

\citet{Sierks_2011} have modeled a global shape of (21) Lutetia, combining
two techniques: stereo-photoclinometry \citep{Gaskell_2008} using images
obtained by OSIRIS, and inversion of a set of 50 photometric light curves and
contours of adaptive optics images \citep{Carry_2010,Kaasalainen_2011}. Twelve 
different shape model solutions are listed in the Planetary Data System
(\href{http://sbn.psi.edu/pds/}{PDS}\footnote{
  \href{http://sbn.psi.edu/pds/}{http://sbn.psi.edu/pds/}}). 

In this work we selected the shape model that has 2962 faces from
the PDS database.  The body is aligned with the principal axes of inertia,
in such a way that the inertia tensor becomes a diagonal matrix. Thus, the
x-axis is aligned with the smallest moment of inertia (longest axis), while
the z-axis is aligned with the largest (shortest axis), and the y-axis is
aligned with the intermediate one. The spin velocity of (21) Lutetia is
assumed to be uniform around its maximum moment of inertia ($z$ axis) with
a period of $8.168270 \pm 0.000001$ hours \citep{Carry_2010}.
The algorithm of \citet{Werner_1997} was used to calculate the spherical
harmonic coefficients $C_{n,m}$ and $S_{n,m}$ up to degree 4
(Table \ref{Table01_Harmonics_Lutetia}), considering a uniform bulk
density of $3.4 ~ g.cm^{-3}$.  Please notice that these coefficients
are presented as a reference for describing the exterior gravitational
potential. They can be used to verify the orientation of our
shape. If we fix the expansion of the gravitational field around the
center of mass, we have C 11 = S 11 = 0, and if the axes are exactly
oriented along the principal axes of inertia, we have C21 = S 21 = S 22 = 0
\citep{Scheeres_2000}. However, we did not use these coefficients in our
analyses, our approach (mascon) employs the shape of the asteroid to calculate
the exterior gravitational potential, which is more accurate than the
harmonic coefficients even if this coefficients were measured up to
a higher degree than four.

The algorithm of \citet{Mirtich_1996} provides these values of
moments of inertia divided by the total mass of the body:

\begin{eqnarray*}
 I_{xx} / M&=&802.929   ~~ km^{2} \nonumber \\ 
 I_{yy} / M&=&1096.555  ~~ km^{2}           \\
 I_{zz} / M&=&1263.996  ~~ km^{2} \nonumber
\end{eqnarray*}

From the moments of inertia, we can solve for the equivalent ellipsoid
according to \citet{Dobrovolskis_1996}. The semi-major axes found
are: $62.402 \times 49.254 \times 39.859$ km

As discussed by \citet{Hu_2004}, the main gravity coefficients are
directly related to the principal moments of inertia (normalized by the
body mass) and the unit is the distance squared.

\begin{eqnarray*}
 C_{20} &=& -\frac{1}{2M}(2I_{zz} - I_{xx} - I_{yy})= -314.254 \,km^2   \\
 C_{22} &=&  \frac{1}{4M}(I_{yy} - I_{xx})= 73.406 \,km^2       \\
\end{eqnarray*}

\begin{figure}[!htp]
   \includegraphics[width=0.48\linewidth]{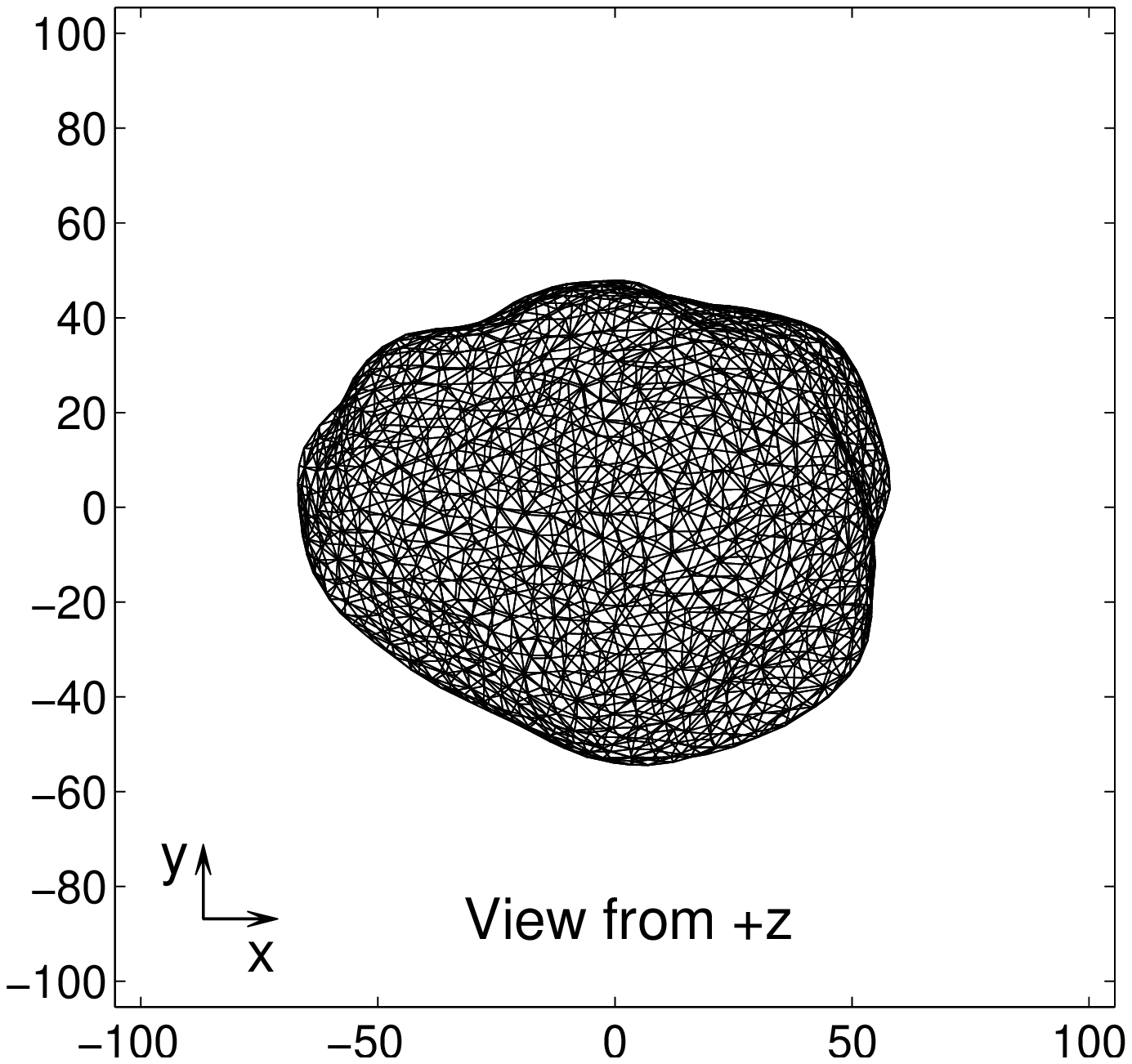}
   \includegraphics[width=0.48\linewidth]{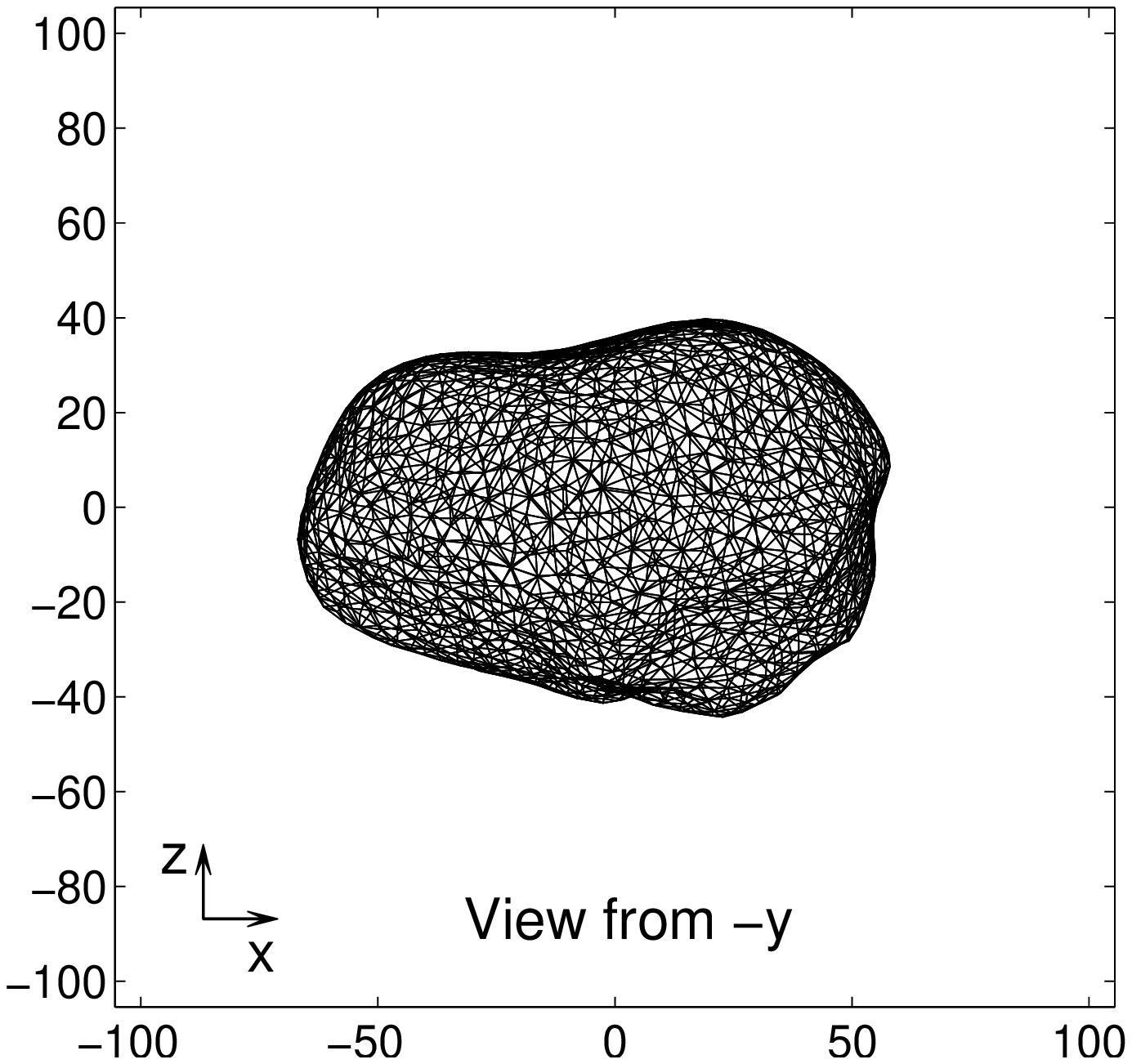}\\
   \vspace{0.45cm}
   \includegraphics[width=0.48\linewidth]{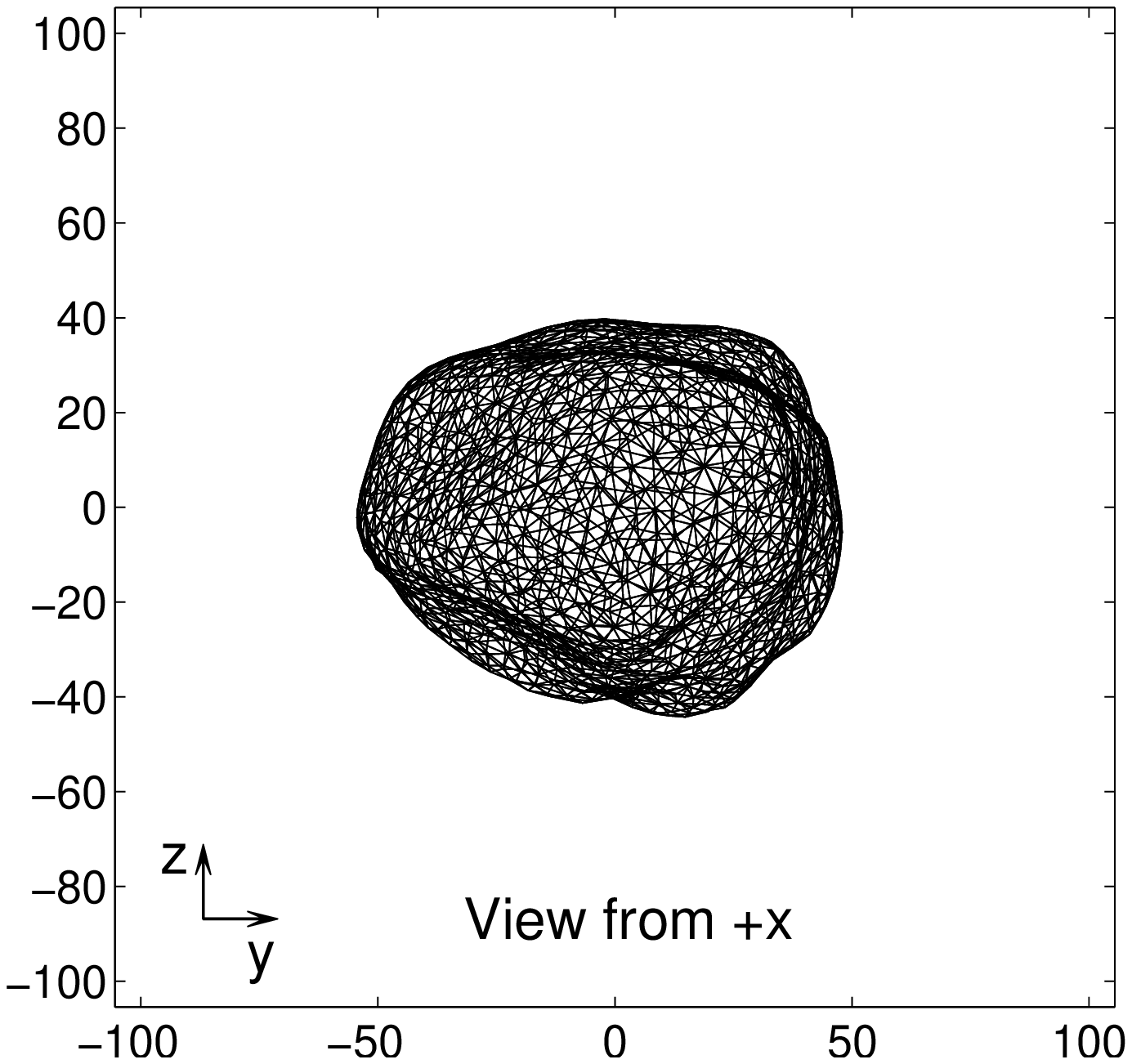}   
   \includegraphics[width=0.48\linewidth]{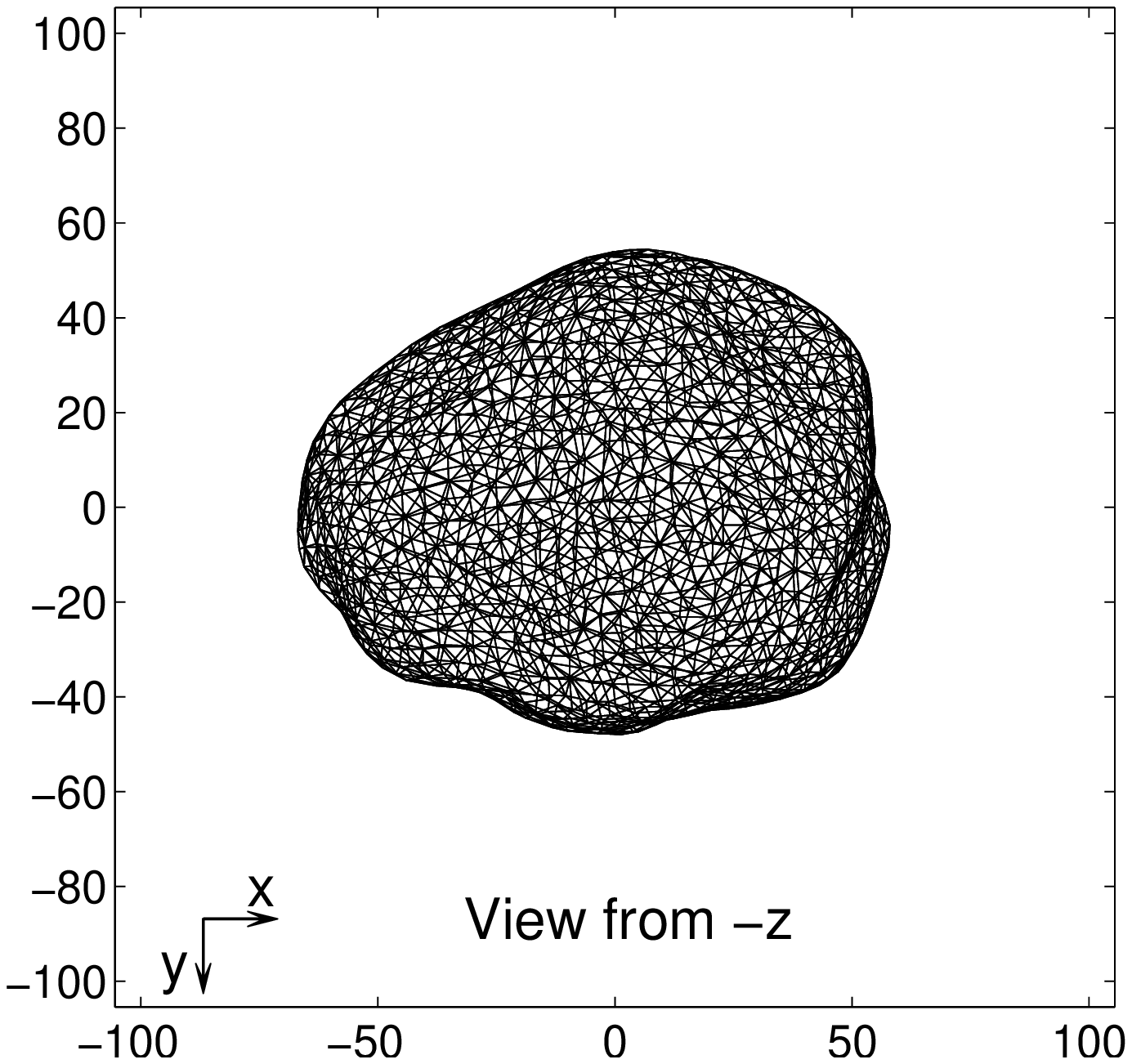}\\
   \hspace{0.15cm}
   \includegraphics[width=0.48\linewidth]{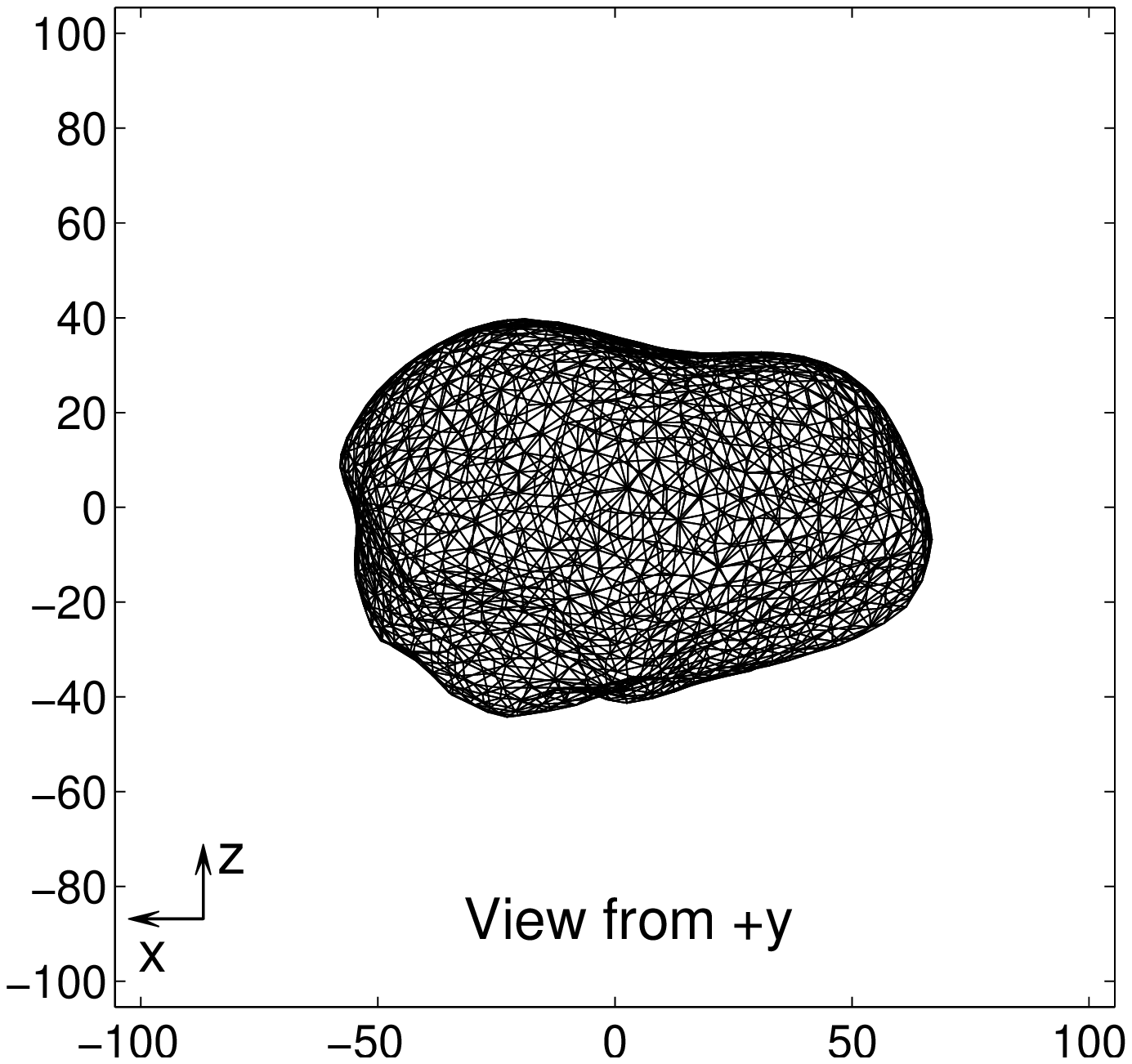}
   \includegraphics[width=0.48\linewidth]{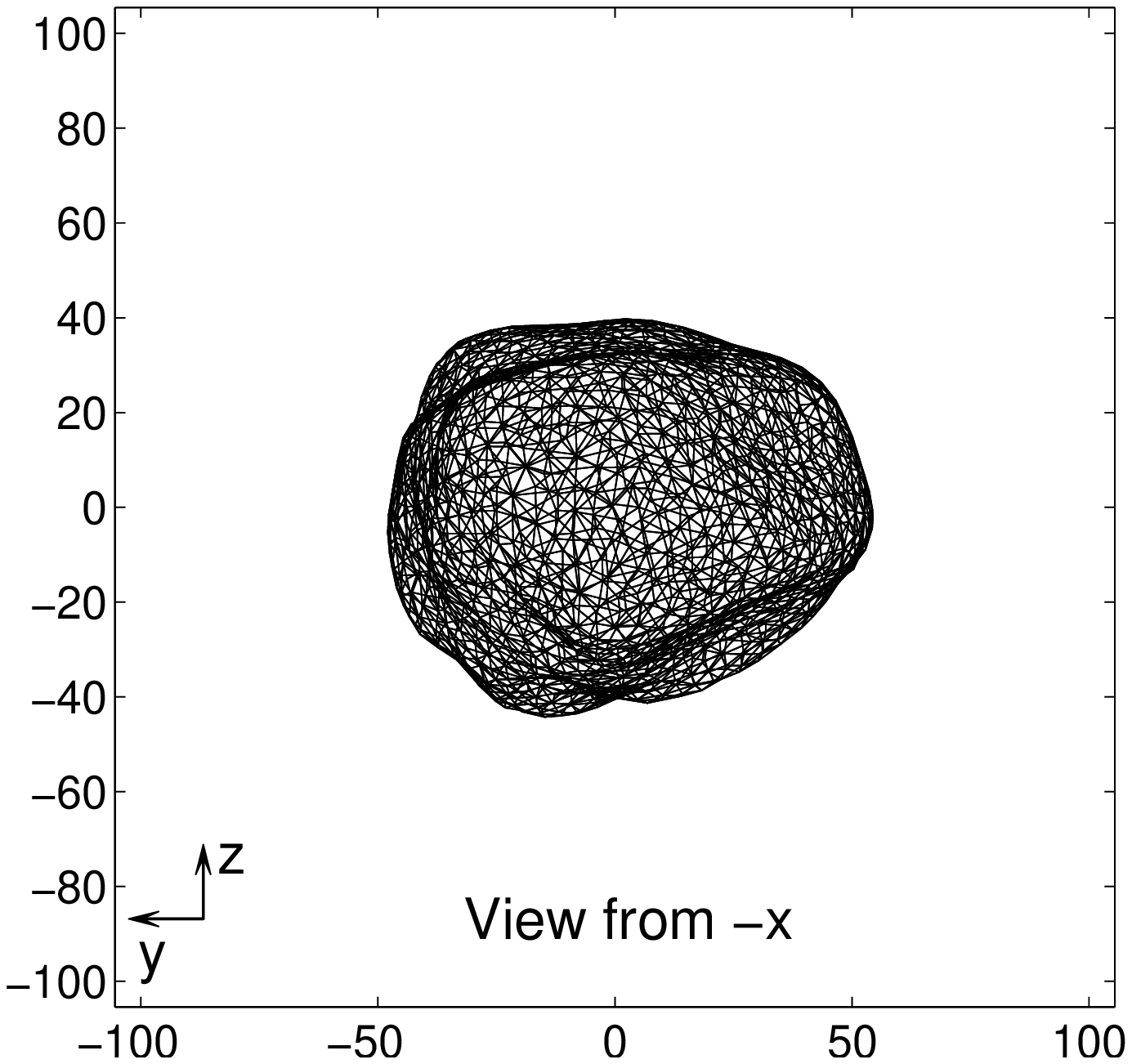}\\
   \caption{Polyhedral shape of (21) Lutetia shown in 6 perspective views ($\pm$ x, $\pm$ y, and $\pm$ z), 
   using the shape model provided by PDS database with 2962 triangular faces \citep{Sierks_2011}, after aligning 
   the asteroid with the principal axes of inertia.}
   \label{Fig03_Shape_Lutetia}
\end{figure}

A mass-distribution parameter $\sigma$ can determined to be:

\begin{eqnarray*}
 \sigma = \frac{I_{yy}-I_{xx}}{I_{zz}-I_{xx}} = -\frac{4C_{22}}{C_{20}-2C_{22}}= 0.637
\end{eqnarray*}

this value of $\sigma$ denotes that Lutetia is not close to the
rotational symmetry about the z-axis ($\sigma=0$) or x-axis ($\sigma=1$).
That clearly appears in the elongated shapes viewed from various
perspectives presented in Fig. \ref{Fig03_Shape_Lutetia},
with overall dimensions ($km$) of $(-66.854, 57.959) \times (-54.395,
47.920) \times (-44.238, 39.721)$ in the x,y, and z directions, respectively,
and a polyhedral volume of $495140.993 ~ km^{3}$
(volume-equivalent diameter of 98.155 $km$).

   {\floatstyle{simpleruletable}\restylefloat{table}
   \begin{table}[!htp]
   \begin{minipage}[h]{1\linewidth}
   \caption{Lutetia Gravity Field Coefficients up to order 4, using the shape model of 2962 faces. These coefficients are computed with respect to a constant density of 3.4 $g.cm^{-3}$, a total mass of 1.68 $ \times 10^{18} $  kg (derived from the polyhedron volume), and a reference distance of 49.1 km.}
   \label{Table01_Harmonics_Lutetia}
   \vspace{0.1cm}
   \resizebox{1.0\textwidth}{!}{
   \rowcolors{1}{grey80}{}
   \begin{tabular}{ccrr}
   \hline
   Order & Degree & $ C_{nm}$ & $ S_{nm}$ \\
   \hline
   
  0   &   0   &   1.0000000000       &   \multicolumn{1}{c}{-} \\ 
  1   &   0   &   -2.4161445414 $ \times 10^{-16} $     &   \multicolumn{1}{c}{-} \\ 
  1   &   1   &   4.4587814343 $ \times 10^{-17} $     &   7.2140102052 $ \times 10^{-17} $  \\ 
  2   &   0   &   -1.3047303671 $ \times 10^{-1} $     &   \multicolumn{1}{c}{-} \\ 
  2   &   1   &   2.0156673639 $ \times 10^{-16} $     &   8.9487336812 $ \times 10^{-17} $  \\ 
  2   &   2   &   3.0477066056 $ \times 10^{-2} $     &   8.6521705057 $ \times 10^{-16} $  \\ 
  3   &   0   &   -8.1225875136 $ \times 10^{-3} $     &   \multicolumn{1}{c}{-} \\ 
  3   &   1   &   1.3607877846 $ \times 10^{-2} $     &   6.4377447088 $ \times 10^{-3} $  \\ 
  3   &   2   &   1.7536608648 $ \times 10^{-5} $     &   -3.1776398240 $ \times 10^{-4} $  \\ 
  3   &   3   &   -2.3473257023 $ \times 10^{-3} $     &   1.5994949238 $ \times 10^{-3} $  \\ 
  4   &   0   &   3.5318181727 $ \times 10^{-2} $     &   \multicolumn{1}{c}{-} \\ 
  4   &   1   &   8.1522038541 $ \times 10^{-4} $     &   -4.7670141468 $ \times 10^{-3} $  \\ 
  4   &   2   &   -2.4926821394 $ \times 10^{-3} $     &   1.3305167431 $ \times 10^{-3} $  \\ 
  4   &   3   &   3.8256764962 $ \times 10^{-5} $     &   4.5914129725 $ \times 10^{-4} $  \\ 

   \hline
   \end{tabular}}
   \end{minipage}
   \end{table}
   }

\section{Internal structure of Lutetia}
\label{Internal_structure}

Because of its high IRAS albedo of $0.208 \pm 0.025$, (21) Lutetia was
classified as M-type asteroid by \citet{Barucci_1987} and
\citet{Tholen_1989}. Analyzing the visible spectrum, \citet{Bus_2002}
classified it as (Xk) on the basis of SMASS II spectroscopic data.
Further spectroscopic observations by 
\citet{Birlan_2004,Barucci_2005,Lazzarin_2004,Lazzarin_2009} suggested a
similarity with the
carbonaceous chondrite spectra that characterize the C-type asteroids.
Analyzing the reflectance spectra, \citet{Busarev_2004} indicated the
possibility of Lutetia being an M-type body covered with irregular layer of
hydrated silicates. The Bus-DeMeo taxonomy of asteroids \citep{DeMeo_2009}
put Lutetia in the Xc subclass. Moreover, the available data from ROSETTA
OSIRIS images has been analyzed by \citet{Magrin_2012} and compared
consistently with ground based observations, but no further deep analysis
was possible, since Rosetta only made a relatively brief observation covering
about 50\% of the surface.

According to \citet{Neumann_2013}, (21) Lutetia may have a differentiated
interior, i.e., an iron-rich core and a silicate mantle. Notice that the
other differentiated asteroid such as (1) Ceres and (4) Vesta have been
visited by a spacecraft (DAWN). Because of its large diameter, we think
that it is reasonable to expect an internal differentiated structure
for (21) Lutetia as well.  To understand the effects that
such differentiation may have on the orbits of probes,
we will study the dynamics in the vicinity of (21) Lutetia examining the
effect of its in-homogeneity, considering two distinct models, based on
a three-layers and a four-layers assumption, respectively, as 
already used for other differentiated objects.

\subsection{The three-layers internal model }

\begin{figure}[!htp]
   \includegraphics[width=0.95\linewidth, height=0.95\linewidth]{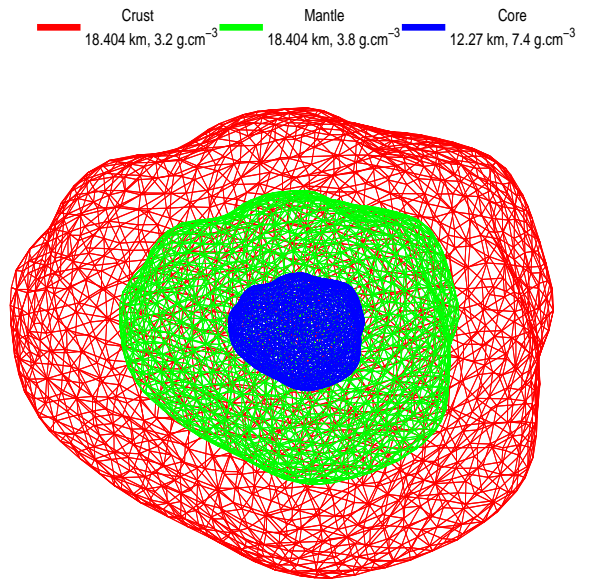}
   \caption{Three-layer structure of (21) Lutetia.}
   \label{Fig02_3layer_Lutetia}
\end{figure}

Our three-layers model is similar to that discussed in
\citet{Park_2014} and \citet{Konopliv_2014}. It corresponds to a
volume-equivalent diameter of 98.155 $km$, in which a crust with a mean
thickness of 18.404 $km$ occupies 75.59\% of the total volume with a
density of 3.2 $g.cm^{-3}$, that represents 71.06\% of the total mass.
The mantle thickness of the asteroid is also modeled with a 18.404 $km$
thickness (22.85\% of the total volume) and a density of 3.8 $g.cm^{-3}$
(25.54\% of the total mass). The core, based on iron meteorites
characteristics, is considered with a 12.27 $km$ thickness (1.56\% of
the total volume) and a density of 7.4  $g.cm^{-3}$ (3.4\% of the total
mass). This structure is exhibited in Figure \ref{Fig02_3layer_Lutetia}. 

\subsection{The four-layers internal model }

\begin{figure}[!htp]
   \includegraphics[width=0.95\linewidth, height=0.95\linewidth]{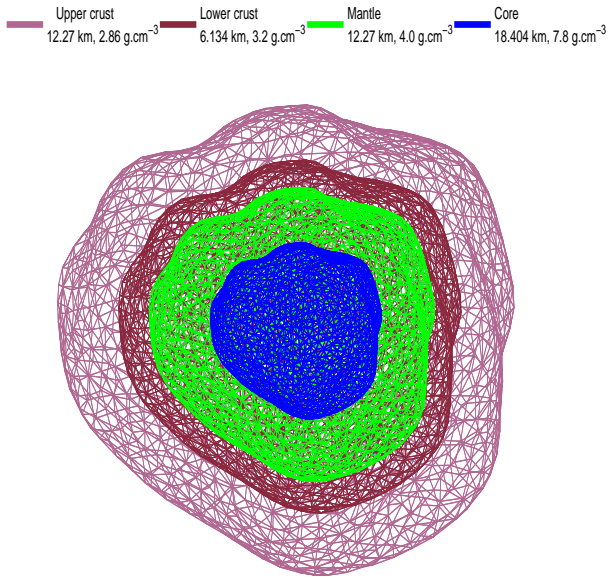}
   \caption{Four-layer structure of (21) Lutetia.}
   \label{Fig03_4layer_Lutetia}
\end{figure}

A more sophisticated model of the internal structure of (21) Lutetia can be
based on the model of Vesta discussed in \citet{Zuber_2011}. It consists
in four layers, shown in Figure \ref{Fig03_4layer_Lutetia}. This model
still includes an iron meteorite core with a thickness of 18.404 $km$
(5.27\% of the total volume) and a density of 7.8 $g.cm^{-3}$ (12.1\%
of the total mass). The mantle thickness is supposed to be 12.27 $km$
(19.15 \% of the total volume) with a density of 4.0 $g.cm^{-3}$ (22.52\%
of the total mass). In that specific model the crust itself is divided
into an upper and lower layers with limits at respectively 12.27 $km$ and
6,13 $km$ thickness. The upper crust represents 57.81\% of the total volume
with a density of 2.86 $g.cm^{-3}$ (48.66\% of the total mass), whereas the
lower crust represents 17.77\% of the total volume with a density of 3.2
$g.cm^{-3}$ (16.72\% of the total mass). 

The two internal structures proposed for (21) Lutetia are
summarized in Table \ref{Table01_internal_structures}. The layers size and
density are constrained to the model of internal structures of Vesta
discussed in \citet{Park_2014, Konopliv_2014} and \citet{Zuber_2011}.
We preserve the total mass of Lutetia by fixing the medium density at
$3.4 ~ g.cm^{-3}$. In other words, the distribution of the gravity of Lutetia
is changed in the three- and four-layers models to be greater at the center, 
while the mean density is the same as in the uniform structure.

   {\floatstyle{simpleruletable}\restylefloat{table}
   \begin{table*}[!htp]
   \begin{minipage}[h]{1\linewidth}
   \caption{Three- and Four-layered structure of (21) Lutetia}
   \label{Table01_internal_structures}
   
   \resizebox{1.0\textwidth}{!}{
   \rowcolors{1}{grey80}{}
   \begin{tabular}{lcccc}
   \hline
               & Thickness  &    Density    &           Volume         &           Mass             \\
               & ($km$)     & ($g.cm^{-3}$) & (\% of the total volume) &  (\% of the total mass)    \\
   \hline
   \multicolumn{5}{c}{Three-layer model} \\
   Core        & 12.270     &     7.40       &            1.56          &              3.40           \\
   Mantle      & 18.404     &     3.80       &           22.85          &             25.54         \\
   Crust       & 18.404     &     3.20       &           75.59          &             71.06         \\
   \hline
   \multicolumn{5}{c}{Four-layer model} \\
   Core        & 18.404    &     7.80       &            5.27          &             12.10          \\
   Mantle      & 12.27     &     4.00       &           19.15          &             22.52          \\
   Lower Crust & 6.13      &     3.20       &           17.77          &             16.72           \\
   Upper Crust & 12.27     &     2.86       &           57.81          &             48.66          \\

   \hline
   \end{tabular}}
   \end{minipage}
   \end{table*}
   }

\subsection{Influence of the internal models on the gravitational potential}
 
For assessing the effects of the two different internal structures above
described on the external potential of (21) Lutetia, we used the shape model
with 2962 triangular faces and applied the approach of
\citet{Chanut_2015a}, dividing each tetrahedron into up to eight parts
(Mascon 8), to at 980396 points placed in an equally spaced grid
generated from the surface of the asteroid up to 200 km in the (x,y) plane.
Mascon 8 seems to be satisfactory in terms of precision and computational
time. Higher divisions could provide somewhat better accuracy but at
a heavier computational cost.

Also, using the shape of the asteroid to model the external
  gravitational field according to the equation 9 in this paper or the
  equation 4 in \citet{Chanut_2015a} is actually
  more accurate. According to \citet{Park_2014}, the spherical
  harmonic series may not converge close the surface but the polyhedral
  approach is guaranteed to converge outside of the polyhedron.
  
\begin{figure}[!ht]
   \includegraphics[width=0.48\linewidth]{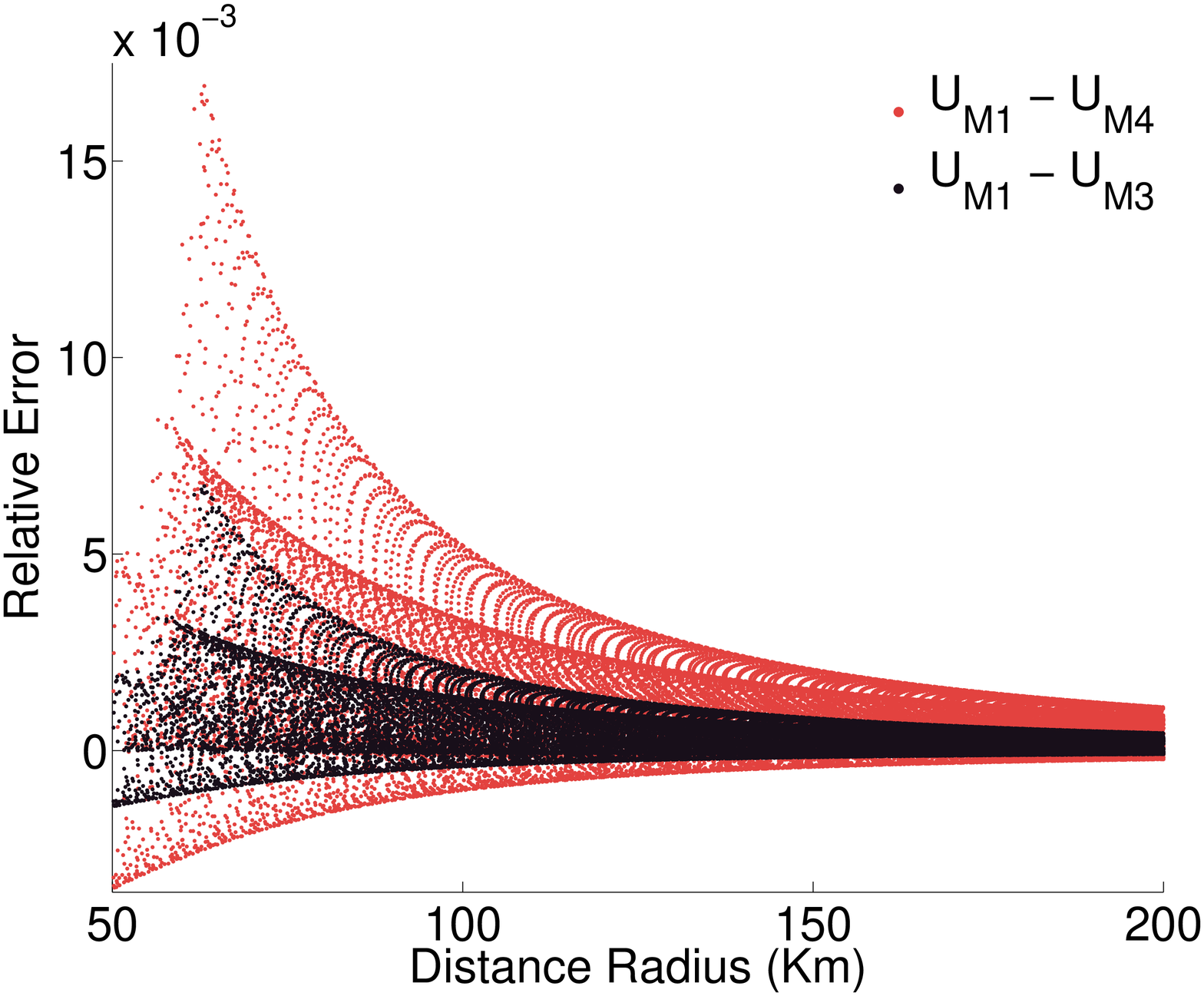}
   \includegraphics[width=0.48\linewidth]{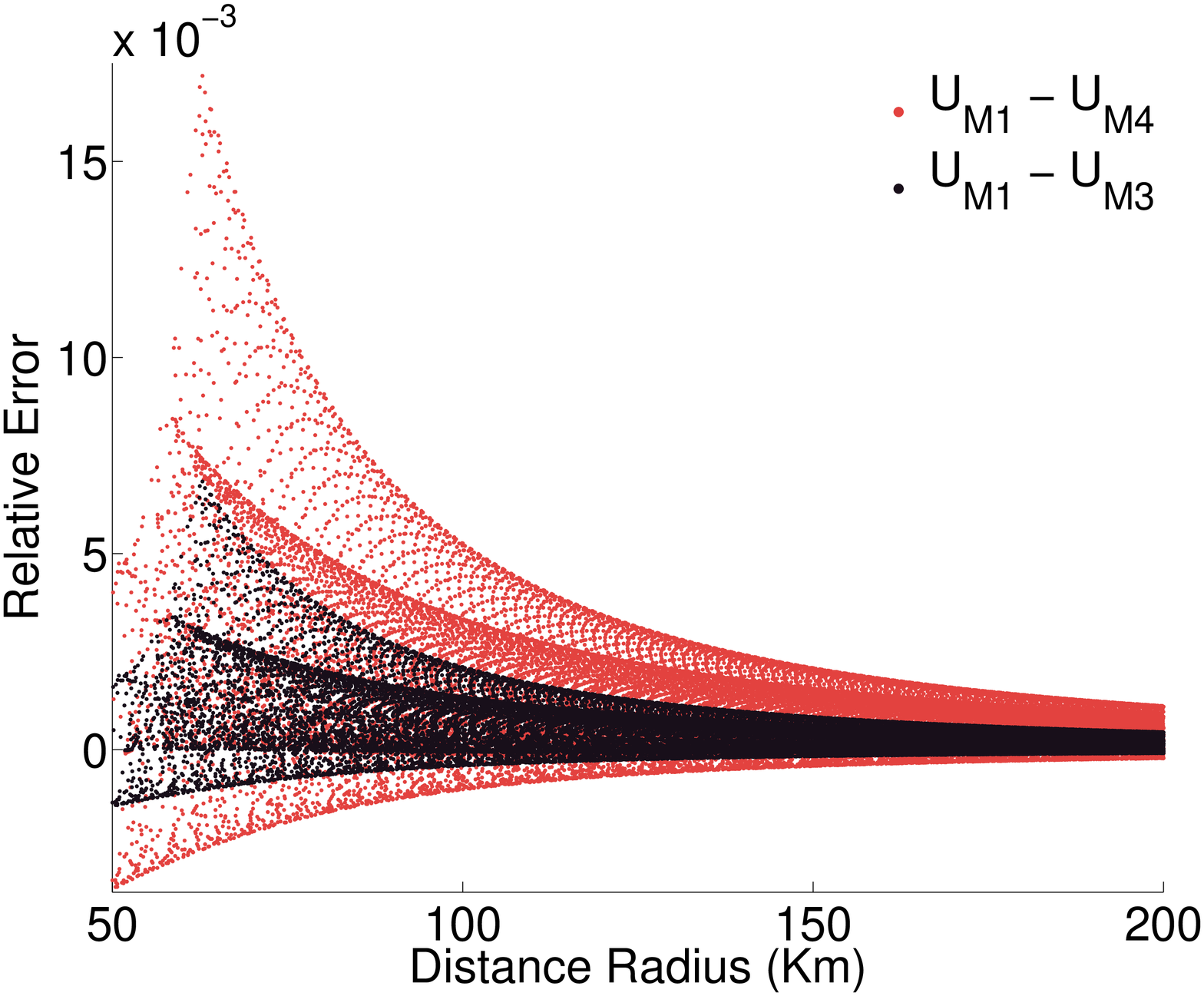}
   \caption{Relative difference of the gravitational potential calculated
     with Mascon 8 considering a uniform density $U_{M1}$ with the model,
     considering a four-layers structure (red dots) or a three-layers
     structure (black dots), using the shape model with 11954 triangular
     faces (left-hand side) and the shape model with 2962 triangular
     faces (right-hand side).}
   \label{Fig04_Relative_error}
\end{figure}

In Fig. \ref{Fig04_Relative_error} (left-hand side) we present the
relative difference of the gravitational potential considering an uniform
density $U_{M1}$ with the four-layers structure (red dots) or the
three-layers structure (black dots). The figure shows that the relative
difference is inversely proportional to the distance from the surface
of the asteroid, and the potential calculated near the surface is affected
significantly by the internal structure.  Moreover, the shape model with
11954 triangular faces is also used in this work to calculate the same
relative difference, and presented in Fig. \ref{Fig04_Relative_error}
(right-hand side). A very good agreement between the two shapes
was found. In terms of CPU time, the total simulation time on a Pentium 3.8 GHz
CPU took about 16 minutes using the first shape model, while required
54 minutes with the second one.  That guided us to use the model with
2962 faces for the rest of this work. 

\section{Equations of motion and dynamical properties in the
  vicinity of Lutetia}
\label{Dynamical_properties}

In this section, we evaluate the dynamical environment close to (21) Lutetia
caused by its in-homogeneous structure, and its consequences on any
spacecraft orbiting around it.  First, we consider a zone where the effect
of the solar gravity is considerably smaller than the asteroid gravity, that
is to say a region inside its Hill sphere. Its Hill radius
$R_{H}= r \sqrt[3]{\frac{m}{3M_{\text{\sun}}}}$ varies between 20042
$km$ at perihelion $(r_p)$ and 27897 $km$ at aphelion $(r_a)$.  As an
example, within 300 km from the asteroid center of mass, the solar gravity
perturbation reaches $1.43 \times 10^{-12} ~ m.s^{-2}$ at perihelion
and $7.38 \times 10^{-13} ~ m.s^{-2}$ at aphelion. That is completely
negligible compared with the total gravitational attraction exercised
by the asteroid on the spacecraft, which is $1.26\times10^{-6} km.s^{-2}$.

Another perturbation that arises from the Sun is the Solar radiation
pressure (SRP). Generally, the magnitude of the SRP acceleration ($g$)
appears in the Hill equation of motion as a linear term in the
first integral \citep{Scheeres_2002}.  Assuming that the spacecraft is a
flat plate oriented to the Sun, the SRP acceleration $g$ always acts in
the anti-solar direction.  It is computed as:

\begin{equation}
  g = \frac{\beta}{d^2}, 
 \end{equation}
 
where $\beta=\frac{(1+\eta)G1}{B}$ is the SRP parameter,
$G1 = 1 \times 10^8 kg.km^3s^{-2}m^{-2}$ is the solar constant
\citep{Giancotti_2014}, $\eta$ is the reflectance of the spacecraft
material (equal to 0 for perfectly absorbing material and to 1
for perfect reflection), $B$ is the spacecraft mass to area ratio in
$kg.m^{-2}$ usually computed by dividing the total mass by the projected
surface area of the spacecraft and $d$ is the heliocentric distance of
the asteroid in $km$. Taking into account the physical characteristics of
a Rosetta-like spacecraft, i.e. a maximum projected area of $65 ~m^{2}$ and
a mass of $1400 ~kg$ \citep{Scheeres_1998a}, the total solar radiation
pressure acceleration  varies from $2.58 \times 10^{-17} ~ m.s^{-2}$ up to
$5.00 \times 10^{-17} ~ m.s^{-2}$ at the aphelion and perihelion distance
from the Sun, respectively.  After considering the above calculations,
we neglect the effects of both the SRP and the solar gravity in our model.

\subsection{Equations of motion}

According to \citet{Scheeres_1996,Scheeres_1999,Scheeres_2012}, in the
absence of any solar perturbations, the equations of motion of a
spacecraft orbiting a uniformly rotating asteroid and significantly far
from any other celestial body are

\begin{eqnarray}
    \ddot{x}-2\omega \dot{y} &=& \omega^2 x +U_{x} \label{Equations_motion1}\\  
    \ddot{y}+2\omega \dot{x} &=& \omega^2 y +U_{y} \label{Equations_motion2}\\ 
    \ddot{z}&=& U_{z}.
    \label{Equations_motion3}
 \end{eqnarray}
where $U_x$, $U_y$ and $U_z$ are the first-order partial
derivatives of the potential $U(x,y,z)$ and $\omega$ is the spin rate of
the asteroid. 

Because Eqs. (\ref{Equations_motion1}) to (\ref{Equations_motion3}) are
time invariant, the Jacobi constant exists as an additional integral
of motion.  The Jacobi integral for the equations of motion is conserved
and is explicitly calculated as
 \begin{eqnarray}
    C=\underbrace{\frac{1}{2} \omega^2(x^2+y^2)+U(x,y,z)}_{\text{Modified potential}} 
    - \underbrace{\frac{1}{2} (\dot{x}^2+\dot{y}^2+\dot{z}^2)}_{\text{Kinetic energy}}
    \label{Jacobi}
 \end{eqnarray}

\subsection{Zero velocity surfaces and equilibria}
 
As shown in Eq. (\ref{Jacobi}), the Jacobi integral is a relation
between the possible position of the particle and the kinetic energy
with respect to the rotating asteroid. If the particle's velocity
becomes zero, the zero-velocity surfaces are defined by

\begin{eqnarray}
    C=\frac{1}{2} \omega^2(x^2+y^2)+U(x,y,z)  = V(x,y,z)
\end{eqnarray}

where $V(x,y,z)$ is the Modified potential.

This equation defines zero-velocity surfaces depending on the asteroid shape
and also on the value of $C$. These surfaces are all evaluated close to the
critical values of $C$ and intersect or close in upon themselves at points
called equilibrium points \citep{Scheeres_1996}. The location of these
equilibrium points can be found by solving the equation
$\nabla V(x,y,z) = 0$. The number of solutions depends on the shape
and on the spin rate of the asteroid. 

\begin{figure}
   \includegraphics[width=0.48\textwidth]{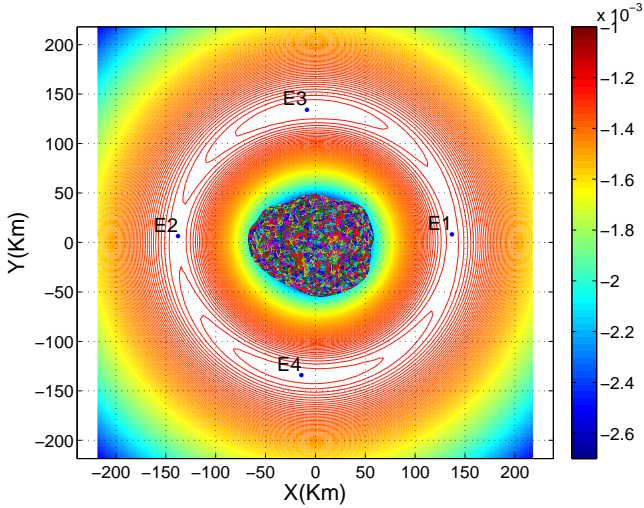}
      
   \caption{Zero-velocity curves and equilibrium points of (21) Lutetia in the
     xoy plane, obtained using the shape model with 2962 triangular faces and
     the four-layers structure.  The colour code gives the intensity of the
     Jacobi constant in $km^{2} s^{−2}$.  The equilibrium points outside the
     body ($E_{1}, E_{2}, E_{3},E_{4}$) are displayed in the figure.}
   \label{Fig05_Zero_velocity}
\end{figure}

Using the shape model of (21) Lutetia with 2962 triangular faces,
the projection of the zero-velocity surface onto the $z=0$ plane is shown
in Figure \ref{Fig05_Zero_velocity}. The zero-velocity curves of the
asteroid have four solutions outside the body, separated by
approximately $90^{\circ}$ in longitude.  Only the external equilibria are
presented in this work since there is not a good agreement between Mascon
8 and the classical polyhedron method inside the body. However, the
agreement is better outside and near the body surface \citep{Chanut_2015a}.

Figs. \ref{Fig05_Zero_velocity} displays the results from the Mascon
8 model and the four-layers structure.  Results for the classical
polyhedral model, for the Mascon 8 gravity model with uniform density,
and for the Mascon 8 with three-layers are similar and will not be displayed,
for brevity. The maximum difference
between the classical polyhedral approach and the Mascon 8 considering a
uniform density occurs at the location of E4 (0.143 km, which
represents 0.11\% of the distance from the center of the body), and
being less than the null hypothesis level, may therefore
be considered satisfactory.  

We observe that the positions of the equilibrium points (E1, E2, E3, E4)
are moved by up to 0.112, 0.137, 0.0614, and 0.143 km considering the 
three-layers structure, and up to 0.294, 0.351, 0.173, and 0.417 km
considering the four-layers one, respectively.

   {\floatstyle{simpleruletable}\restylefloat{table}
   \begin{table*}[!htp]
   \begin{minipage}[h]{1\linewidth}
   \caption{Locations of equilibrium points of (21) Lutetia and their Jacobi constant C (using the shape model of 2962 faces), generated by the classical polyhedral model \citep{Tsoulis_2001} and the Mascon 8 gravity model \citep{Chanut_2015a}.} \label{Table03_Equilibrium_Points_1483}
   
   \resizebox{1.0\textwidth}{!}{
   \rowcolors{1}{grey80}{}
   \begin{tabular}{crrrl}
   \hline
    & \multicolumn{1}{c}{x (km)} & \multicolumn{1}{c}{y (km)} & \multicolumn{1}{c}{z (km)} & \multicolumn{1}{c}{$C(km^{2} s^{-2}$)}\\
   \hline

   \multicolumn{5}{c}{Polyhedral model, uniform density} \\ \hline
   E1 & 137.10784172 & 8.44279347 & 0.08555291 & -0.12634256 $ \times 10^{-2} $   \\
   E2 & -138.19144378 & 6.56551358 & 0.04185644 & -0.12679936 $ \times 10^{-2} $   \\
   E3 & -8.70389476 & 134.01690523 & 0.03696436 & -0.12441019 $ \times 10^{-2} $   \\
   E4 & -14.61831274 & -134.06107222 & 0.08749509 & -0.12467120 $ \times 10^{-2} $   \\   
   \multicolumn{5}{c}{Mascon 8, uniform density} \\ \hline
   E1 & 137.08769819 & 8.38960864 & 0.09077155 & -0.12632687 $ \times 10^{-2} $   \\
   E2 & -138.15024726 & 6.54070646 & 0.04488269 & -0.12677495 $ \times 10^{-2} $   \\
   E3 & -8.66458628 & 134.02811212 & 0.03854483 & -0.12441517 $ \times 10^{-2} $   \\
   E4 & -14.47666720 & -134.07545174 & 0.09370561 & -0.12467157 $ \times 10^{-2} $   \\ 
   \multicolumn{5}{c}{Mascon 8, three-layered structure} \\ \hline
   E1 & 136.99825452 & 8.32304070 & 0.08727020 & -0.12626501 $ \times 10^{-2} $   \\
   E2 & -138.01547466 & 6.50802298 & 0.04322724 & -0.12669215 $ \times 10^{-2} $   \\
   E3 & -8.61381286 & 134.06254558 & 0.03698190 & -0.12443281 $ \times 10^{-2} $   \\
   E4 & -14.33458820 & -134.09482376 & 0.08998671 & -0.12467656 $ \times 10^{-2} $   \\ 
   \multicolumn{5}{c}{Mascon 8, four-layered structure} \\ \hline
   E1 & 136.86707031 & 8.19520730 & 0.08068557 & -0.12617328 $ \times 10^{-2} $   \\
   E2 & -137.81293855 & 6.44518383 & 0.04046285 & -0.12656817 $ \times 10^{-2} $   \\
   E3 & -8.51569913 & 134.11522533 & 0.03433857 & -0.12445911 $ \times 10^{-2} $   \\
   E4 & -14.06158951 & -134.12920961 & 0.08318837 & -0.12468367 $ \times 10^{-2} $   \\ 
 
   \hline
   \end{tabular}}
   \end{minipage}
   \end{table*}
   }

As reported in many previous studies \citep{Szebehely_1967, Scheeres_1994, Murray_1999, Hu_2008, Yu_2012, Jiang_2014, Wang_2014}, we can also examine
the stability of the equilibria determined above. The linearized state
equations in the neighborhood of the equilibrium points are summarized as 

\begin{eqnarray}
  \ddot{X} -2\omega\dot{Y} + U_{xx}X + U_{xy}Y + U_{xz}Z &=& 0\nonumber\\
  \ddot{Y} +2\omega\dot{X} + U_{xy}X + U_{yy}Y + U_{xz}Z &=& 0  \label{state_equations}\\
  \ddot{Z} + U_{xz}X + U_{xz}Y + U_{zz}Z &=& 0   \nonumber
\end{eqnarray}

where $X=x-x_{L}$, $Y=y-y_{L}$,$Z=z-z_{L}$, and ($x_{L}$,$y_{L}$,$z_{L}$)
denote the coordinates of the equilibrium point,
$U_{\zeta\eta}=\frac{\partial^{2} U}{\partial \zeta \partial\eta}:
\zeta, \eta= x,y,z$. The eigenvalues of the equation \ref{state_equations} are
calculated by finding the roots of the characteristic equation at the
equilibrium point. 

\begin{eqnarray}
  & &\lambda^{6} + \alpha\lambda^{4}+\beta\lambda^{2}+\gamma=0
\end{eqnarray}

where $\lambda$ is the eigenvalues, $\alpha=U_{xx} + U_{yy} + U_{zz} + 4\omega^{2}$, $\beta =U_{xx} U_{yy}+ U_{yy} U_{zz} + U_{zz} U_{xx} - U_{xy}^{2}- U_{yz}^{2} - U_{xz}^{2}+ 4\omega^{2} U_{zz}$, $\gamma=U_{xx} U_{yy} U_{zz} + 2U_{xy} U_{yz} U_{xz} - U_{xx} U_{yz}^{2} - U_{yy} U_{xz}^{2} - U_{zz} U_{xy}^{2}$.
For more information, we recommend that interested readers review equation 14
in \citet{Jiang_2014}. The linearization method is applied using the
classical polyhedral model and the Mascon 8 approach with uniform density
and Mascon 8 gravity model considering the two multiple layers structures.
This requires calculating the second derivatives of the potential that
results in correcting the analytical form already presented in
\citet{Chanut_2015a} with the following expression:

\begin{eqnarray}
U = \sum\limits_{i=1}^{n} \frac{\mu_{i}}{r_{i}} &\Rightarrow& U_{\zeta}= \sum\limits_{i=1}^{n} -\frac{\mu_{i}\zeta_{i}}{r_{i}^{3}} \nonumber \\
\frac{\partial }{\partial \eta} (U_{\zeta})&=& \sum\limits_{i=1}^{n}\bigg[-\frac{\mu_{i}}{r_{i}^{3}} \frac{\partial }{\partial \eta} (\zeta_{i}) +\frac{3\mu_{i} \zeta_{i}\eta_{i}}{r_{i}^{5}}\bigg]
\end{eqnarray}
Eq. (7) leads to the second derivatives
\begin{eqnarray}
U_{\zeta\zeta}&=& \sum\limits_{i=1}^{n} \bigg[-\frac{\mu_{i}}{r_{i}^{3}}+\frac{3\mu_{i} \zeta_{i}^{2}}{r_{i}^{5}}\bigg]\nonumber \\ 
U_{\zeta\eta}&=& \sum\limits_{i=1}^{n} \bigg[\frac{3\mu_{i} \zeta_{i}\eta_{i}}{r_{i}^{5}}\bigg]\\
\zeta, \eta&=& x,y,z\nonumber \text{    and } \zeta \neq \eta
\end{eqnarray}

where $r = \sqrt{x^{2} + y^{2} + z^{2}}$ represents the distance between the
center of mass of each tetrahedron shaping the asteroid and the external
point. The eigenvalues of the linearized system are listed in
table \ref{Table05_eigenvalues}. The classification of the equilibrium
points defined in \citet{Jiang_2014,Wang_2014} shows that E1 and E2
belong to Case 2 (two pairs of imaginary eigenvalues and one pair of real
eigenvalues).  As a consequence, the saddle  equilibrium points are
unstable, whereas E3 and E4 belong to Case 1 (purely imaginary
eigenvalues), that leads to a linear stability of center equilibrium
points. Thus, according to the classification originally proposed
by \citet{Scheeres_1994}, (21) Lutetia can be classified as a Type I
asteroid  . We can conclude that the effects of the two layer structures
chosen on the stability of the equilibria are not determining.

   {\floatstyle{simpleruletable}\restylefloat{table}
   \begin{table*}[!htp]
   \begin{minipage}[h]{1\linewidth}
   \caption{Eigenvalues of the coefficient matrix at the 4 external equilibrium points.} \label{Table05_eigenvalues}
   
   \resizebox{1.0\textwidth}{!}{
   \rowcolors{1}{grey80}{}
   \begin{tabular}{crlrlrlrl}
   \hline
   
   Eigenvalues & \multicolumn{2}{c}{E1} & \multicolumn{2}{c}{E2} & \multicolumn{2}{c}{E3} & \multicolumn{2}{c}{E4} \\ \hline
   
 \multicolumn{9}{c}{\citet{Tsoulis_2001} considering the uniform density} \\

    & \multicolumn{2}{c}{$ \times 10^{-3}$} & \multicolumn{2}{c}{$ 10^{\times -3}$} & \multicolumn{2}{c}{$ 10^{\times -3}$} & \multicolumn{2}{c}{$ \times 10^{-3}$} \\ \hline 
$\lambda_{1}$ &       0.220401 & $i$ &       0.225039 & $i$ &       0.215898 & $i$ &       0.217647 & $i$ \\ 
$\lambda_{2}$ &      -0.220401 & $i$ &      -0.225039 & $i$ &      -0.215898 & $i$ &      -0.217647 & $i$ \\ 
$\lambda_{3}$ &       0.217889 & $i$ &       0.223367 & $i$ &       0.186893 & $i$ &       0.195130 & $i$ \\ 
$\lambda_{4}$ &      -0.217889 & $i$ &      -0.223367 & $i$ &      -0.186893 & $i$ &      -0.195130 & $i$ \\ 
$\lambda_{5}$ &       -0.068854  &   &       -0.096044  &   &       0.098844 & $i$ &       0.076583 & $i$ \\ 
$\lambda_{6}$ &        0.068854  &   &        0.096044  &   &      -0.098844 & $i$ &      -0.076583 & $i$ \\ 
\hline

    \multicolumn{9}{c}{Mascon 8 considering the uniform densities} \\ 

    & \multicolumn{2}{c}{$ \times 10^{-3}$} & \multicolumn{2}{c}{$ 10^{\times -3}$} & \multicolumn{2}{c}{$ 10^{\times -3}$} & \multicolumn{2}{c}{$ \times 10^{-3}$} \\ \hline 
$\lambda_{1}$ &       0.220329 & $i$ &       0.224875 & $i$ &       0.215878 & $i$ &       0.217595 & $i$ \\ 
$\lambda_{2}$ &      -0.220329 & $i$ &      -0.224875 & $i$ &      -0.215878 & $i$ &      -0.217595 & $i$ \\ 
$\lambda_{3}$ &       0.217873 & $i$ &       0.223245 & $i$ &       0.187309 & $i$ &       0.195345 & $i$ \\ 
$\lambda_{4}$ &      -0.217873 & $i$ &      -0.223245 & $i$ &      -0.187309 & $i$ &      -0.195345 & $i$ \\ 
$\lambda_{5}$ &       -0.068571  &   &       -0.095370  &   &       0.098100 & $i$ &       0.076186 & $i$ \\ 
$\lambda_{6}$ &        0.068571  &   &        0.095370  &   &      -0.098100 & $i$ &      -0.076186 & $i$ \\ 
\hline

 \multicolumn{9}{c}{Mascon 8 considering the three-layered structure} \\

    & \multicolumn{2}{c}{$ \times 10^{-3}$} & \multicolumn{2}{c}{$ 10^{\times -3}$} & \multicolumn{2}{c}{$ 10^{\times -3}$} & \multicolumn{2}{c}{$ \times 10^{-3}$} \\ \hline 
$\lambda_{1}$ &       0.220074 & $i$ &       0.224448 & $i$ &       0.215787 & $i$ &       0.217422 & $i$ \\ 
$\lambda_{2}$ &      -0.220074 & $i$ &      -0.224448 & $i$ &      -0.215787 & $i$ &      -0.217422 & $i$ \\ 
$\lambda_{3}$ &       0.217726 & $i$ &       0.222875 & $i$ &       0.188941 & $i$ &       0.196268 & $i$ \\ 
$\lambda_{4}$ &      -0.217726 & $i$ &      -0.222875 & $i$ &      -0.188941 & $i$ &      -0.196268 & $i$ \\ 
$\lambda_{5}$ &       -0.067270  &   &       -0.093478  &   &       0.095128 & $i$ &       0.074283 & $i$ \\ 
$\lambda_{6}$ &        0.067270  &   &        0.093478  &   &      -0.095128 & $i$ &      -0.074283 & $i$ \\ 
\hline

 \multicolumn{9}{c}{Mascon 8 considering the four-layered structure} \\

    & \multicolumn{2}{c}{$ \times 10^{-3}$} & \multicolumn{2}{c}{$ 10^{\times -3}$} & \multicolumn{2}{c}{$ 10^{\times -3}$} & \multicolumn{2}{c}{$ \times 10^{-3}$} \\ \hline 
$\lambda_{1}$ &       0.219691 & $i$ &       0.223786 & $i$ &       0.215654 & $i$ &       0.217165 & $i$ \\ 
$\lambda_{2}$ &      -0.219691 & $i$ &      -0.223786 & $i$ &      -0.215654 & $i$ &      -0.217165 & $i$ \\ 
$\lambda_{3}$ &       0.217510 & $i$ &       0.222310 & $i$ &       0.191206 & $i$ &       0.197591 & $i$ \\ 
$\lambda_{4}$ &      -0.217510 & $i$ &      -0.222310 & $i$ &      -0.191206 & $i$ &      -0.197591 & $i$ \\ 
$\lambda_{5}$ &       -0.065291  &   &       -0.090502  &   &       0.090804 & $i$ &       0.071479 & $i$ \\ 
$\lambda_{6}$ &        0.065291  &   &        0.090502  &   &      -0.090804 & $i$ &      -0.071479 & $i$ \\ 
\hline

   \hline
   \end{tabular}}
   \end{minipage}
   \end{table*}
   }

\section{Orbital stability about Lutetia}
\label{Orbital_stability}

The goal of this section is to evaluate what should be the influence of
Lutetia internal structure on the trajectory of a spacecraft in a close
orbit. Thus we numerically investigate the perturbations on
initially equatorial orbits.  In particular, we focus our analysis
on the effects of the layered structures on limiting stability against
impacts, so as to help us choosing the limits for periapsis radius in
our stability analysis.


\subsection{Stability Against Impact}
According to \citep{Scheeres_2000, Chanut_2014, Chanut_2015b}, the
stability against impact is devoted to characterize the spacecraft
dynamics, choosing initial conditions in such a way that the spacecraft
stays in the outer portion of the zero-velocity curve, and the value
of the Jacobi integral is smaller or equal to a specific value,
corresponding to the minimum value of the Jacobi constant
at the equilibrium point E2 listed in
Table \ref{Table03_Equilibrium_Points_1483}.  A simple check in terms of
osculating orbital elements (periapsis radius, eccentricity, and initial
longitude) for an equatorial orbit is applied in order to determine the
occurrence of an impact with the surface:

\begin{eqnarray}
  \frac{-\mu (1+e)}{2r_{p}}+W\sqrt{\mu r_{p}(1+e)} + U(r=r_{p},\lambda) + J_{0} = 0
\end{eqnarray}

\begin{figure}[!ht]
   \includegraphics[width=0.95\linewidth]{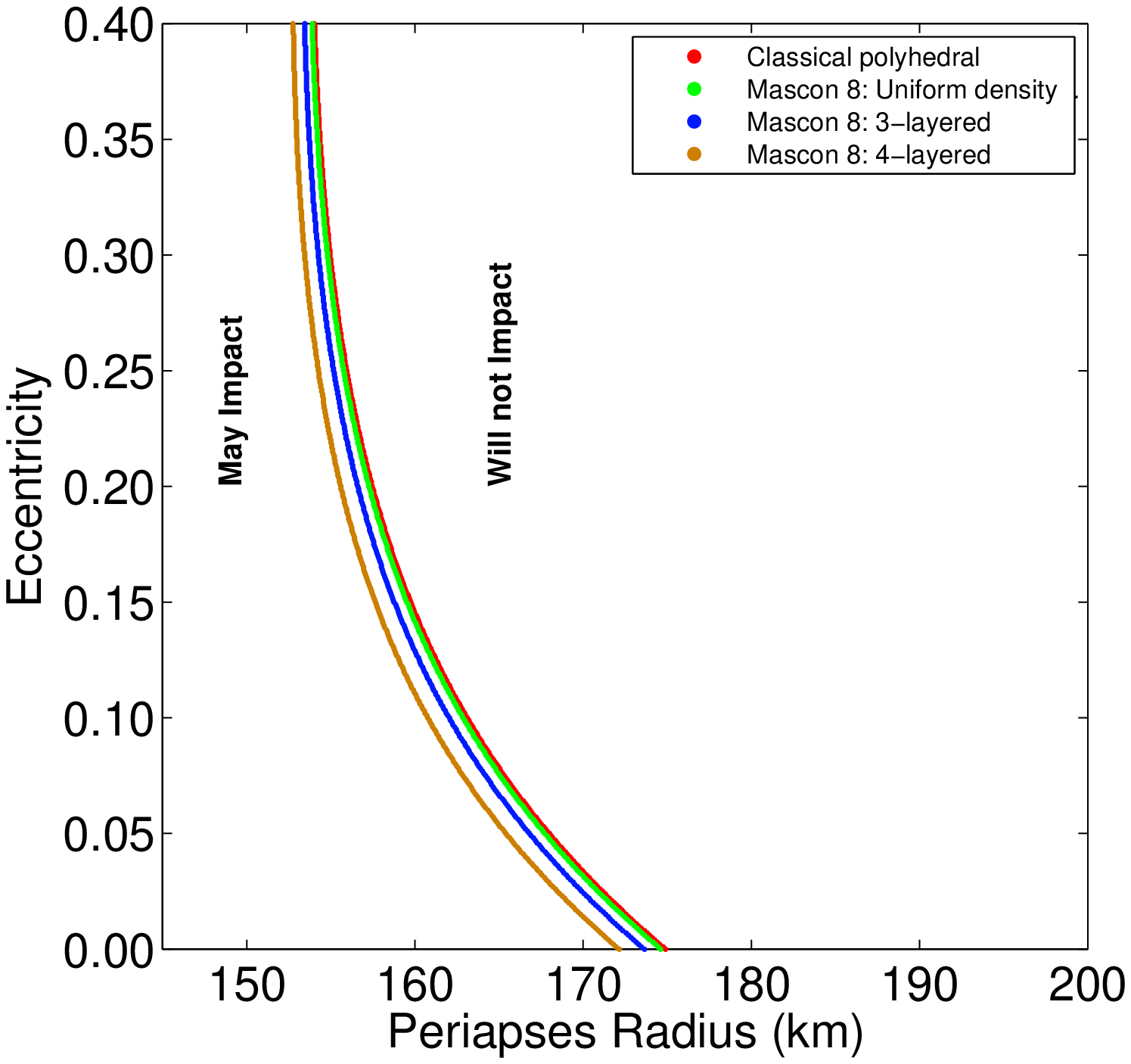}      
   \caption{Stability against impact curve for equatorial, direct orbits around (21) Lutetia. The colours correspond to different approaches for calculating the potential : a classical polyhedral approach considering a uniform density and a Mascon 8 gravity model assuming the three- and four-layers structures shown in Table \ref{Table01_internal_structures}.} \label{Fig06_Stability_Impact}
\end{figure}

According to this last equation, the limits for the zones of stability
against impact for Lutetia  are shown in Figure
\ref{Fig06_Stability_Impact}. The initial orbits that do not undergo impact
with Lutetia correspond to the right-hand side of the curves. We remark that
the curves related to the classical polyhedral approach and the Mascon 8
gravity model assuming a uniform density can hardly be distinguished. On the
contrary, the curves related to multiple layers structures are quite distinct
and get closer to the surface which means that the non impacting zones are
larger than in the case of uniform density. The eccentricities in this
analysis are limited to 0.4 because orbits with high eccentricities have a
small perihelion distance, which implies that
the spacecraft will travel at a high relative velocity when encountering
the asteroid, and that would make these orbits unpractical
for a space mission.  We also notice that the initial
eccentricity is not a primary parameter in affecting the periapsis
distance: the periapsis distance is moved from a little less than 160 km
for $e = 0.4$ to a little more than 175 km for $e = 0$. Thus,
studying the stability against impact leads to conclude that orbits must
lie outside of 175 km from Lutetia to avoid an impact on the surface. 

\subsection{Stability analysis}

In this section we present a numerical survey performed to find 
stable orbits around (21) Lutetia, with a period of 45 days, corresponding
to more than 70 orbits around the asteroid. For this purpose, we
consider the three different models of its internal structure. This work
concentrates mainly on equatorial and prograde orbits. An orbit is
considered stable if the oscillations of its eccentricity do not
exceed a threshold value, although the orientation of these orbits
may change. Thus, our task consists in observing the oscillation of
$e(t)$ around its initial value. However, an alternative way for finding
a stable orbits could be to measure oscillation in the periapsis radius
instead of the the oscillation of $e(t)$, that could be enhanced in future work.
Following the previous section, orbits with
a periapsis distance ($r_p$) between 150 and 200 km from the asteroid center
with an interval of 2 km are tested using the Bulirsch-Stoer integrator.
We consider initially circular ($e_{ini}=0$) or slightly eccentric orbits
(with initial eccentricity of respectively 0.05, 0.1 and 0.2).  For
the sake of simplicity, initial conditions are chosen in such a way that
each test particle is at the periapsis distance on the equatorial plane of
the body ($i =0$), with 12 different longitudes $\lambda$ varying from
$0^{\circ}$ to $330^{\circ}$. Even with this discrete grid, a through
exploration of the three-dimensional initial phase space
($r_p$, $e$, $\lambda$) requires 26 (periapsis radius) $\times$ 4
(eccentricities) $\times$ 12 (longitudes) = 1248 initial conditions for
each model of the internal structure of Lutetia.  The initial conditions in
inertial space calculated from the two-body problem in the body-fixed
reference frame are:

\begin{eqnarray*}\label{initial_conditions}
  x = r_{p} \cos{\lambda} & & \dot{x} = \textcolor{white}{-}\bigg[\sqrt{\frac{\mu}{r}(1+e)} - r \omega\bigg]\sin \lambda  \\  
  y = r_{p} \sin{\lambda} & & \dot{y} = -\bigg[\sqrt{\frac{\mu}{r}(1+e)} - r \omega\bigg]\cos \lambda   \\  
  z = 0  \;\;\;\;\;\;\;\; \;\;                & & \dot{z} = 0  \ 
\end{eqnarray*}

The orbital position and velocity calculated in the rotating frame can then
be transformed into position and velocity in the inertial frame with a simple
approach. As already mentioned in the previous section, the new Mascon 8
approach, implemented by \citet{Chanut_2015a}, is chosen to calculate the
gravitational field of the equations of motions in Eqs.
(\ref{Equations_motion1}),(\ref{Equations_motion2}),
and (\ref{Equations_motion3}).

\begin{figure*}[!htp]
   \centering    
   \includegraphics[width=0.32\linewidth]{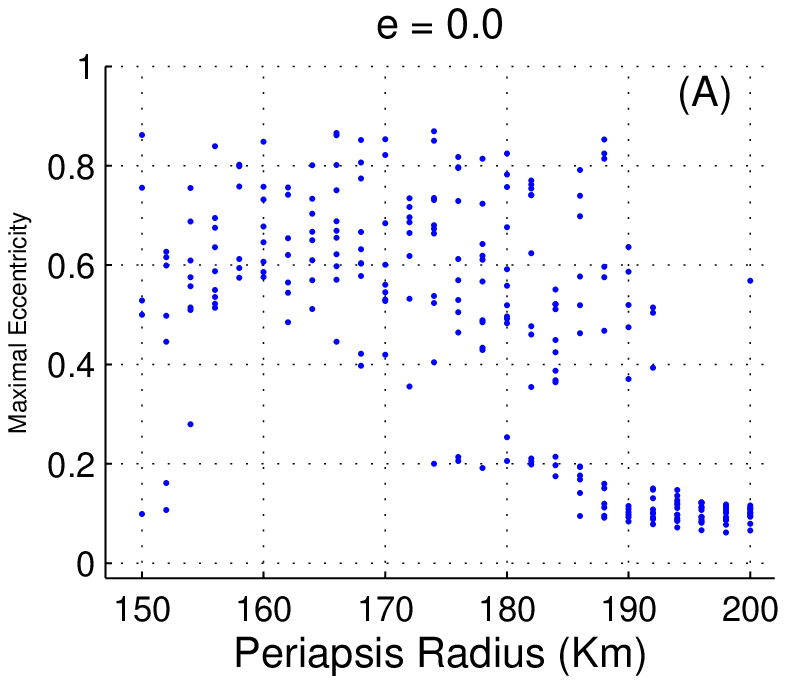}
   \includegraphics[width=0.32\linewidth]{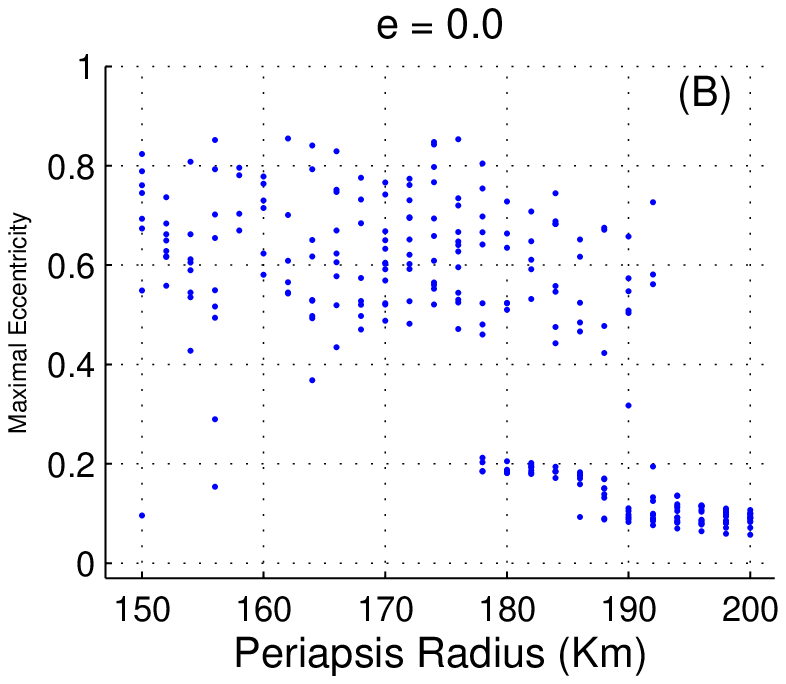}
   \includegraphics[width=0.32\linewidth]{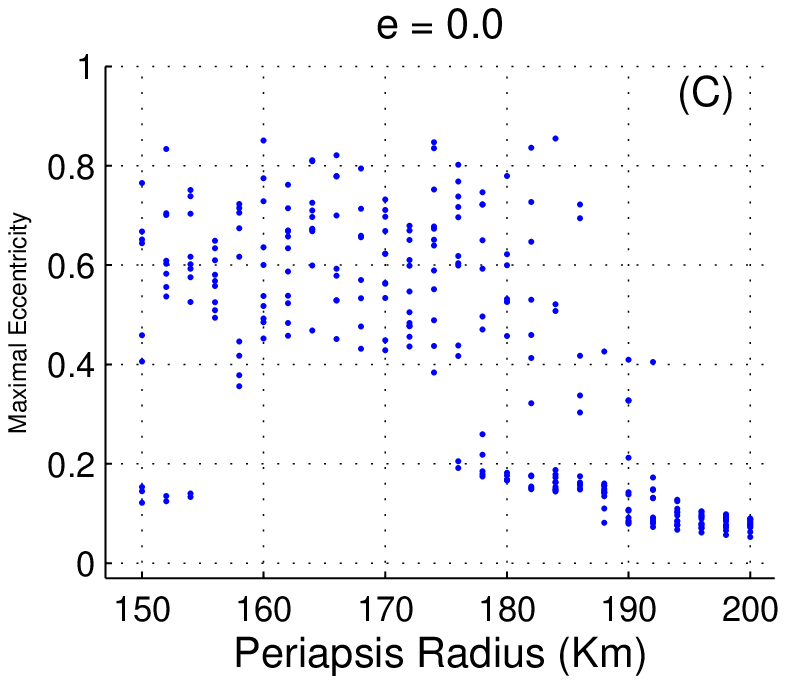} \\
   \caption{The maximal eccentricity of initially circular orbits about (21) Lutetia after 45 days, considering the one- (A), three- (B) and four-layer structure (C).} \label{Fig07_Max_ecc_circular_orbits}
\end{figure*}

After eliminating orbits colliding with the body\footnote{As a first
  approximation, an ellipsoid with semimajor axes of
  $62.402 \times 49.254  \times 39.859$ km is considered for detecting
  collisions. A more detailed study of every collisional 
event is, in our opinion, beyond the scope of this work}, Fig. 
\ref{Fig07_Max_ecc_circular_orbits} shows the maximum eccentricities of
initially circular orbits, after 45 days, considering the uniform
(Fig. \ref{Fig07_Max_ecc_circular_orbits}a), three-layers (Fig.
\ref{Fig07_Max_ecc_circular_orbits}b) and four-layers (Fig. 
\ref{Fig07_Max_ecc_circular_orbits}c) structure of Lutetia. As a further
general comment, one can notice that, within the considered area, no orbit
escapes from the system. Nevertheless, a large majority of orbits suffer
strong perturbations due to the irregular structure of Lutetia. An 
example of this behaviour can be seen in the three
panels of Fig.~\ref{Fig07_Max_ecc_circular_orbits}.
Objects starting with perfectly circular orbits experience changes
in eccentricity of 0.06 after 45 days. These changes are not large
enough to affect the stability of an eventual probe over the mission
period, but could potentially be hazardous for longer timescales.
Interested readers could find more information on results
for orbits with larger initial eccentricities ($e_{ini}>0$) in appendix \textcolor{blue}{1}.

Despite similarities among different panels of Fig.
\ref{Fig07_Max_ecc_circular_orbits} and
\textcolor{blue}{8}, a simple comparison of the tree panels (a,b,c) for each 
initial eccentricity shows that different internal structures of the asteroid 
could stabilize or destabilize some orbits.  For initially eccentric orbits,
(\textcolor{blue}{8}) shows an increase in the stability region when the
initial eccentricity increases.  Most important, for all the eccentricities
here considered, the stability region increases when the three- and
four-layers structures are considered.

Finally, in order to show the effects of the suggested core-mantle structure
of (21) Lutetia on the orbital stability, three examples of 3D equatorial orbits
after 45 days, are displayed in Figs. \textcolor{blue}{9},
\textcolor{blue}{10} and \textcolor{blue}{11}. The core-mantle
structure can cause orbits to precess or regress around the asteroid,
depending on the initial conditions. In the first examples
(Fig. \textcolor{blue}{9}), considering the uniform structure
destabilize the orbit, while the orbit is destabilized considering
three-layer structure in the second example. Finally, the four-layer
structure stabilizes the orbit of the Fig. \textcolor{blue}{11}

\section{Conclusion}
\label{Conclusion}

The computations carried out in this paper were performed based
on the suggestion that the asteroid (21) Lutetia, the European
space agency's Rosetta mission target, may have an in-homogeneous density.
This lead to the problem of modeling its gravity field considering
three different kinds of internal structures (uniform, three-layers and
four-layers). Our different models of Lutetia structure
were obtained within the Mascon gravity framework,
using the shaped polyhedral source, and dividing each tetrahedron into
eight equal layers. The shape of Lutetia is presented, viewed from various
perspectives after aligning the asteroid with the principal axes of inertia.
The harmonic coefficients $C_{n,m}$ and $S_{n,m}$ up to degree 4, considering
an uniform bulk density, were computed with respect to the reference radius.
Then, two different internal structures for (21) Lutetia were considered
to study the orbital dynamics in its vicinity and to examine the effect of
the in-homogeneity. Both three-layer and four-layer Lutetia models provided
important effects on the external potential. In their study of the gravity
field of Vesta, \citet{Park_2014} have shown that the thin crust model is
the more appropriate representation of Vesta's internal structure.
In our case, the two layer models provide a satisfactory estimation of the
gravitational potential, within 150-200 km from the asteroid center of mass,
with a  maximum relative difference from the uniform density equal
to $9.38\times10^{-4}$. In terms of CPU time requirements, both of the models
are somewhat comparable. However, a better close approach of a spacecraft is
necessary to fit the real gravity data to found the plausible internal
structure of this asteroid.

Correcting the analytical form of the second derivatives of the potential
presented in \citep{Chanut_2015a}, we tested the stability of the equilibria
points. We showed that the location of the equilibrium points can be slightly
changed by up to 0.351 km. Moreover, the limiting planar figure of the
stability against impact gets closer to the body considering the four-layers
structure. Finally, in order to examine the potential effects of the
in-homogeneity of Lutetia, stability analyses were investigated by testing
orbits in an appropriate grid of initial conditions.
Generally speaking , the stability region
increases when considering the three- and four-layers structure. 
Future applications of this model could involve the study of the
stability of polar orbits, that are more suitable for mapping and
reconnaissance purposes.

\section*{Acknowledgments} 

We are grateful to an anonymous referee for comments and suggestions
that greatly improved the quality of this work.
The authors wish to thank the S\~ao Paulo State Science Foundation (FAPESP),
which supported this work via the grants 13/15357-1, 14/06762-2,
16/04476-8, and CNPq (grants 150360/2015-0, and 312313/2014-4).

\label{lastpage}

\end{document}